\documentclass[UTF8,a4paper]{article}
\usepackage{amsmath}
\usepackage{graphicx}
\usepackage{subfigure}
\usepackage{epstopdf}
\usepackage{geometry}
\usepackage{ulem}
\usepackage{indentfirst}
\usepackage{amsfonts,amssymb,dsfont}
\usepackage{setspace}
\usepackage{threeparttable}
\usepackage{amsmath,mathtools,amsthm}
\usepackage{array}
\usepackage{extarrows}
\usepackage{upgreek}

\makeatletter
\renewcommand*\env@matrix[1][\arraystretch]{%
  \edef\arraystretch{#1}%
  \hskip -\arraycolsep
  \let\@ifnextchar\new@ifnextchar
  \array{*\c@MaxMatrixCols c}}
\makeatother
\begin{document}

%QED 3+1: Appelquist T W, Bowick M, Karabali D and Wijewardhana L C R 1986 Phys. Rev. D 33 3704
%Optical evidence for a Weyl semimetal state in pyrochlore Eu 2 Ir 2 O 7
%Collective Modes of the Massless Dirac Plasma
%relaxation time:Dc and ac transport in silicene
%Femtosecond carrier dynamics and saturable absorption in graphene

\title{\bf Two-dimensional parabolic Dirac system in the presence of non-magnetic and magnetic impurities}
\author{Chen-Huan Wu
\thanks{chenhuanwu1@gmail.com}
%\\Key Laboratory of Atomic $\&$ Molecular Physics and Functional Materials of Gansu Province,
\\College of Physics and Electronic Engineering, Northwest Normal University, Lanzhou 730070, China}

\maketitle
\vspace{-30pt}
\begin{abstract}
\begin{large} 

We theoretically investigate the effect of the non-magnetic and magnetic impurities to the 2D parabolic Dirac system.
The induced charge density by the charged impurity is obtained by the linear response theory within the random phase
approximation.
We also detailly calculate the RKKY interaction between two magnetic impurities placed within the 2D sheet of the Dirac materials
with isotropic and anisotropic dispersion.
For the anisotropic dispersion,
the RKKY interaction is also anisotropic and related to the lattice parameters which can be obtained throuh the DFT calculation or
the experiments.
The fearures of the RKKY interaction also can be treated as a signature of the topological phase transition
as well as the change of Berry curvature.
Our results are also illuminating to the study of the static screening and the RKKY interaction of the isotropic or
anisotropic 3D Dirac or Weyl systems.\\
\\
{\bf PACS number(s)}: 73.20.At, 67.85.De\\

\end{large}

\end{abstract}
\begin{large}

\section{Introduction}

For topological insulator with the spin-momentum locking,
the low-energy effective model near the critical point of the phase-transition (from band insulator to semimetal) 
exhibits rich physical properties.
It's also found that the ferromagnetic phase transition occur in 2D system which exhibits
a sharp van Hove singularity (VHS) near the Fermi level and the out-of-plane spin-polarization 
 by controlling the hole doping (e.g., through the gate voltage),
like the monolayer GaSe which is in a hexagonal structure with two sublayers\cite{Cao T},
or the monolayer MoS$_{2}$ which is also in a hexagonal structure with three sublayers\cite{Cheng Y C}.
The VHSs locate at the points where the Fermi surface touch the brillouin zone boundary.
The large density of states (DOS) at VHSs also leads to instability and compressibility
as well as the higher probability for the phase-transition,
and even the nontrivial response to the metal, like the Anderson's orthogonality catastrophe\cite{Gefen Y}.
The umklapp scattering between the vicinity of the VHSs also leads to the spin susceptibility\cite{Honerkamp C}.
%{Instabilities of interacting electrons on the triangular lattice}
The famous Friedel oscillation or the RKKY interaction (oscillation) can also be treated as a fingerprints of the
topological phase-transition in the presence of the single-impurity (independent of the momentum),
as found in the silicene or germanene\cite{Chang1,Zare M,Duan H J}:
the beating of the friedel oscillation vanishes for gapless (semimetal) silicene,
while for the magnetic impurities localed at the zigzag-edge of silicene,
the RKKY interaction vanishes once the edge model enters into the band insulator from the topological insulator.
That also indicates the nontrivial topology of the materials.
In topological materials,
 the quantum critical point seperates the electron metal and the hole metal which 
with Fermi energy $E_{F}$ larger or smaller than zero, respectively,
 can be simply controlled by the out-of-plane electric field
or in-plane magnetic field\cite{Dominguez F},
but note that in the absence of out-of-plane magnetic field,
the charged-density-wave (CDW; the spin-independent staggered potential term)
order won't appear even at low-temperature\cite{Yang K Y}.
 Then the case of long-range Coulomb interaction which couples to the relativistic fermions at low-temperature,
the critical temperature for the CDW phase transition
(for the case of small gap with $k_{B}T\gg E_{F}$) won't be charged\cite{Yang K Y}.

In this article, we focus on the parabolic 2D electronic system in the presence of the non-magnetic and magnetic impurities.
In the presence of non-magnetic charged impurity,
its long-range Coulomb interaction is statically screened by the conduction electron within the sheet of the 2D Dirac material.
For Thomas-Fermi wave vector $q_{TF}>2k_{F}$ where $k_{F}$ is the Fermi wave vector,
the stromg screening approximation\cite{Adam S} is valid except for the case of high carrier density ($\gg 10^{-12}$ cm$^{-2}$).
In the systems with long-range Coulomb interaction,
the method of random-phase-approximation (RPA) as an infinite diagrammatic resummation\cite{Sabio J}
can be use to explore the density-density response function as well as the current-current response function
in the presence of the dynamical screening by the conduction electron,
and it also leads to the collective excitation like the plasmon modes as well
as the dynamical structure factor within the fluctuation-dissipation theorem\cite{Reed J P,Sabio J}
with the charge response function at finite temperture.
While for the magnetic impurities placed within the plane of the 2D Dirac system,
the local magnetic moment is also screened by the kondo effect with the carrier density larger than the impurity density\cite{Ruhman J}.
The indirect exchange interaction between two magnetic impurities is mediated by the RKKY interaction through the itinerant electrons.
In this case, for the topological insulators which with spin-momentum locking,
an anisotropic DM term\cite{Ye F} would appears in the absence of the inversion symmetry.

It's been proved\cite{Wang S X,Shiranzaei M} that the Rahsba-coupling, which is linear with the momentum,
has an non-negligible impact on the RKKY
interaction between magnetic impurities in the 3D Dirac/Weyl as well as the noncentrosymmetric systems.
It also found that the Rashba-coupling affects deeply the Ising and Dzyaloshinskii-Moriya (DM) interaction
in both of the one-dimension (1D), 2D and 3D system\cite{Klinovaja J,Imamura H}.
Although the Rashba-coupling breaks the symmetry of (real) spin space by mixing the spin-up and spin-down components,
it won't affects the separation of the pseudospin components,
and that provides us a way to form the retarded real-space Green's function as we done in this article.
For the isotropic 2D Dirac system like the intrinsic silicene,
the isotropic band dispersion leads to an isotropic RKKY interaction,
while for the anisotropic 2D Dirac system,
like the black phosphorus,
whose in-plane coupling and Fermi velocity are unequal in $x-$ and $y-$direction,
RKKY interaction is related to the alignment of the two magnetic impurities.

\section{Screened Coulomb potential}

In the presence of conduction electron, the screened Coulomb potential
reads 
\begin{equation} 
\begin{aligned}
V^{s}_{q}&=e^{2}\int\frac{e^{i(q_{x}x+q_{y}y)}e^{-k_{s}|{\rm R}|}}{4\pi\epsilon\epsilon^{*}|{\rm R}|}dxdy\\
&=\frac{e^{2}e^{-z\sqrt{q^{2}+k_{s}^{2}}}}{2\epsilon\epsilon^{*}\sqrt{q^{2}+k_{s}^{2}}},
\end{aligned}
\end{equation}
where $q=\sqrt{q_{x}^{2}+q_{y}^{2}}=2k{\rm sin}(\theta/2)$ is the scattering wave vector
with the scattering angle $\theta$ 
(note that the nonzero $q$ leads to the charge-nonneutrality), 
%{Quantum Hall effects in a Weyl semimetal: Possible application in pyrochlore iridates}
$|{\rm R}|=\sqrt{x^{2}+y^{2}+z^{2}}$ is the distance between charged impurity (within the substrate) to the conduction electron,
$k_{s}=2\pi e^{2}\Pi(q,\omega)/\epsilon\epsilon^{*}$ is the screening wave vector (or the inversal screening length).
The screened long-range Coulomb potential here is in contrast with the on-site Hubbard interaction as well as the on-site Kondo interaction.
 While for the strong-screening approximation\cite{Adam S},
e.g., when the Thomas-Fermi wave vector $q_{TF}\gg q$
when means that the Coulomb potential as well as the Wigner-Seitz radiu are very large,
the Coulomb potential reads $V^{os}_{q}=\frac{2\pi e^{2}}{\epsilon\epsilon^{*}q_{TF}}$.
 %{Boltzmann transport and residual conductivity in bilayer graphene}
However, in the ultraviolet limit where $\sqrt{\omega}\gg q$,
the Coulomb interaction is not efficiently screened\cite{Nandkishore R}.
For the charged impurity which statically screened by the conduction electron,
the induced charge density during one-loop process in static RPA model reads
%{Plasmon-pole approximation for many-body-effects in extrinsic graphene}
\begin{equation} 
\begin{aligned}
\delta n_{r}=&{\rm Ze}\int\frac{d^{2}q}{(2\pi)^{2}}e^{i{\bf q}\cdot{\bf r}}(1-\epsilon(q))\\
=&{\rm Ze}\int\frac{d^{2}q}{(2\pi)^{2}}e^{i{\bf q}\cdot{\bf r}}V_{q}\Pi(q),
\end{aligned}
\end{equation}
where $\epsilon(q)$ is the dielectric function
and $V^{0}_{q}=2\pi e^{2}/\hbar v_{F}\epsilon\epsilon^{*}q$ is the unscreened Coulomb potential (non-interacting).
Note that here we use the RPA and don't consider the many-electron correction
(including the sefl-energy and vertex correction).
After a straighforward compution,
we arrive
\begin{equation} 
\begin{aligned}
%(2*T*Ze*e^2*pi*(D/(2*mu) - 
%(asin(((hv^2*q^2)/(4*D^2 + hv^2*q^2))^(1/2))*(D^2 - (hv^2*q^2)/4))/(hv*mu*q))*(x*log(y + (x^2 + y^2)^(1/2)) - y + y*log(x + (x^2 + y^2)^(1/2))))/jie
\delta n_{r}=&
% \frac{2{\rm Ze} e^{2}\pi}{4\pi^{2}\epsilon\epsilon^{*}}
% (& \frac{-2g_{s}g_{v}e^{2}\mu}{\epsilon\epsilon^{*}\hbar^{2}v_{F}^{2}} 
%(\frac{m_{D}}{2\mu} - 
%\frac{{\rm asin}(((\hbar^{2} v_{F}^{2} q^{2})/(4m_{D}^{2} 
%+ \hbar^{2} v_{F}^{2} q^{2}))^{1/2})(m_{D}^{2} - (\hbar^{2} v_{F}^{2} q^{2} )/4)}
%{\hbar^{2} v_{F}^{2} \mu q})\\
%&\times(q_{x} {\rm ln}
%(q_{y} + (q_{x}^{2} + q_{y}^{2})^(1/2)) - q_{y} + q_{y} {\rm ln}(q_{x} + (q_{x}^{2} + q_{y}^{2})^(1/2)))),\ {\rm for}\ \mu<m_{D},\\
%\delta n_{r}= 
% \frac{-2g_{s}g_{v}e^{2}\mu}{\epsilon\epsilon^{*}\hbar^{2}v_{F}^{2}} 
% \frac{2{\rm Ze}e^{2}\pi(q_{x}{\rm ln}(q_{y} + (q_{x}^2 + q_{y}^2)^(1/2)) 
%- q_{y} + q_{y}{\rm ln}(q_{x} + (q_{x}^2 + q_{y}^2)^(1/2)))}{4\pi^{2}\epsilon\epsilon^{*}}
%,\ {\rm for}\ \mu>m_{D},\ q<2k_{F}\\
%\delta n_{r}= 
%,\ {\rm for}\ \mu>m_{D},\ q>2k_{F}\\
\frac{{\rm Ze}e^{2}}{2\epsilon\epsilon^{*}}
\frac{\sqrt[4]{-1}\sqrt{\pi}\sqrt{q}(16m_{D}^{2}r^{2}-1){\rm erfi}(\sqrt[4]{-1}\sqrt{qr})+2i\sqrt{r}e^{iqr}(q+8im_{D}^{2}r)}{8\sqrt{2\pi}r^{3/2}\sqrt{iqr}}
\bigg|_{\Lambda},\ {\rm for}\ q\gg m_{D},\\
\delta n_{r}=&
%\frac{{\rm Ze}e^{2}}{2\pi\epsilon\epsilon^{*}}\frac{e^{iqr}(qr+i)}{6\pi^{2}mr^{3}}\bigg|_{\Lambda},
\frac{{\rm Ze}e^{2}}{2\epsilon\epsilon^{*}}
\frac{(-1)^{3/4}3\sqrt{\pi q}{\rm erfi}(\sqrt[4]{-1}\sqrt{qr})+2q\sqrt{r}e^{iqr}(3-2iqr)}{12\sqrt{2}\pi^{3/2}m_{D}r^{5/2}\sqrt{iqr}}\bigg|_{\Lambda},
\ {\rm for}\ q\ll m_{D},\\
\end{aligned}
\end{equation}
%here $Ei(z)$ is the exponential integral,
here ${\rm erfi(z)}$ is the imaginary error function,
$\Lambda$ is the curoff in the ultraviolet limit where the integral convergent (see Appendix.A for detail derivation).
We present the exact result of the induced charge density in the Fig.1,
where the imaginary error function is also considered.
We find that, in the case of $q\ll m_{D}$, the induced charge density by the charged impurity decrease rapidly with the increase of distance $r$.
We also make a comparasion with the $r^{-5}$,
and it turns out that our result is consistent with the 3D electronic system
which with a our-of-palne spatial component as reported in Ref.\cite{Yang B J}.
While for the case of $q\gg m_{D}$, 
it increase in an oscillating way, and the amplitude of the oscillation is enhanced with the increase of $q$.

The screening of the anisotropic Fermions
is isotropic for scattering wave vector $q<2k_{F}$,
which is the usual case at zero temperature.
For the charge density induced linearly by the perturbation,
like the self-consistent Kohn-Sham potential\cite{Baroni S},
%{Intrinsic electrical transport properties of monolayer silicene and MoS2 from first principles}
its divergence as well as the kink of the polarization emerge at $q=2k_{F}$ (nesting wave vector),
which can be easily seen in the gapped 2D Dirac system and the traditional 3D system\cite{Lv M}
(though it's supressed due to the volume effect).
%{Phonons and related crystal properties from density-functional perturbation theory}
At the nesting wave vector of the Fermi surface, the spin susceptibility is divergent
while it won't shows any indications for the non-nested Fermi surface\cite{Honerkamp C}.
For non-magnetic impurity, we only discuss the case of isotropic dispersion here,
the dynamical screening in this case can be well studied by using the RPA
in the presence of the long-range divergent Coulomb interaction.
The RPA is still a good method in the case of low-temperature or highly doped\cite{Hwang E H}
even though the many-electron effect is been ignored. 
However, for the case of anisotropic dispersion,
the resulting anisotropic screening is dependent on the \cite{Yan J}
i.e., the anisotropic band structure (or the dispersion) leads to the screening beyond the conventional Thomas-Fermi screening\cite{Yang B J},
while in the conventional Thomas-Fermi screening, the Fermion excitations are coupled to the
short-range (not the on-site Hubbard type) repulsive interaction and forms the Fermi liquid.
%{Quantum criticality of topological phase transitions in three-dimensional interacting electronic systems}

%described by relativistic Dirac fermions coupled to the long-range Coulomb interaction; 

\section{RKKY interaction}

For the topological materials, like the 2D Dirac materials or the 
topologically protected gapless surface state,
%{Dielectric response and novel electromagnetic modes in three-dimensional Dirac semimetal films
where the inversion symmetry or time-reversal symmetry (TRS) can be broken,
e.g., the inversion sysmmetry can be broken by the out-of-plane electric field or the induced
Rashba-coupling while the TRS can be broken by the off-resonance circularly polarized light, magnetic doping,
%{Ruderman-Kittel-Kasuya-Yosida interaction in Weyl semimetals}
 or by the
competition between the Zeeman coupling and Rashba-coupling\cite{Meijer F E},
the Berry gauge field (Berry curvature) of the Bloch electron $\Omega(k)={\bf k}/k^{3}$\cite{Wu C HX} also plays as an important role
in the absence of the inversion symmetry,
and acts as the magnetic monopole carrying the topological charge in Fermi surface\cite{EZ,Wang S X},
the monopole charge also
affects the RKKY interaction.
The broken 
inversion symmetry and TRS are also common in the 3D Dirac or Weyl semimetal\cite{Soluyanov A A} and even the 3DEG\cite{Wang S X},
which provide the source or sink of the Berry curvature\cite{Schoenemann R}.
As we mentioned at the begining,
the RKKY interaction as well as the Friedel oscillation may indicate the nontrivial topology of the materials,
like the topological phase transition,
where the Pontryagin index can be reduced to the spin/valley or charge Cherm number by summing over the Berry curvature in momentum space.
%{Topological phase transition and electrically tunable diamagnetism in silicene}

In 3D Weyl semimetal or some 2D anisotropic materials (with unequal effective-mass in different directions), like the  
MoS$_{2}$ or the black phosphorus,
the emerging anisotropic fermions in critical point, 
show both relativistic and Newtonian dynamics\cite{Yang B J}.
The
anisotropic Dzyaloshinskii-Moriya (DM) term induced by the Ruderman-Kittel-Kasuya-Yosida (RKKY) interaction exist
in the absence of spin-ratotion symmetry and inversion symmetry.
 The anisotropy in 3D Dirac or Weyl semimetal can be measured by 
ratio between the out-of-plane velocity
and the in-plane one due to the significant effect of the momentum $k_{z}$ expecially for the type-II Weyl semimetals\cite{Soluyanov A A},
while for the 2D Dirac system, we use the different effective masses in $x$ and $y$ directions.

The effect of the spin(or pseudopsin)-anisotropic can be detected by studying the RKKY interaction.
Nextly we study the RKKY interaction between two magnetic impurities in the anisotropic 2D Dirac system.
The RKKY interaction is indeed the second-order perturbation of the standard $s-d$ interaction
with an interaction strength much weaker than the bandwidth which is around three times of the intralayer hopping of the 2D Dirac system, 
like the graphene or silicene.
The magnetic behaviors of the silicene and the phosphorene can be induced by the non-magnetic adatoms and the atomic defects\cite{Zheng H,Zare M2}.
The indirect RKKY exchange is mediated by the interacting with the spin
of the conduction electrons of the host materials (including the metal or semiconductor).
We pick the parabolic system as the example, like the bilayer silicene and the black phosphorus
which have an anisotropy electronic structure\cite{Padilha J E}
unlike the intrinsic graphene.

The low-energy Dirac effective model of the bilayer silicene reads\cite{OO,Wu C H1}
\begin{equation} 
\begin{aligned}
%-\eta\hbar v_{F}^{2}\frac{\mathcal{A}^{2}\Omega}{t^{'2}}
%+\eta\hbar v_{F}^{2}\frac{\mathcal{A}^{2}\Omega}{t^{'2}}
H=&\eta\begin{pmatrix}m_{D}^{\eta ++}&\hbar v_{w}(k_{x}+ik_{y})&0&\hbar v_{F}(k_{x}-ik_{y})\\
 \hbar v_{w}(k_{x}-ik_{y})    &m_{D}^{\eta +-}&\hbar v_{F}(k_{x}+ik_{y})&0\\
 0&\hbar v_{F}(k_{x}-ik_{y})   &m_{D}^{\eta -+}&\eta t'\\        
 \hbar v_{F}(k_{x}+ik_{y})   &0&\eta t'&m_{D}^{\eta --}\\  
 \end{pmatrix}\\
&\times
\begin{pmatrix}
0&0&ia\lambda_{R_{2}}(k_{x}-ik_{y})&0\\
 0&0&-i\lambda_{R_{1}}&-ia\lambda_{R_{2}}(k_{x}-ik_{y})\\
-ia\lambda_{R_{2}}(k_{x}+ik_{y})&i\lambda_{R_{1}}&0&0\\
0&ia\lambda_{R_{2}}(k_{x}+ik_{y})    &0&0
 \end{pmatrix},
\end{aligned}
\end{equation}
where $v_{w}=\sqrt{3}at_{w}/2\hbar\sim v_{F}/10\sim 5.5\times 10^{4}m/s$ is the velocity associates with the trigonal warping,
$t_{w}<t_{\perp}$ is the interlayer hopping which gives rise to the trigonal warping.
$v_{F}=5.5\times 10^{5}$ m/s is the Fermi velocity of the freestnding silicene and it can also as large as $v_{F}=1.2\sim 1.3\times 10^{6}$ m/s 
in an Ag-substrate\cite{Tao L}.
The above expression can also be written as
\begin{equation} 
\begin{aligned}
H=&\hbar v_{F}(\eta\tau_{x}k_{x}+\tau_{y}k_{y})+\eta\lambda_{{\rm SOC}}\tau_{z}\sigma_{z}+a\lambda_{R_{2}}\eta\tau_{z}(k_{y}\sigma_{x}-k_{x}\sigma_{y})\\
&-\frac{\overline{\Delta}}{2}E_{\perp}\tau_{z}+\frac{\lambda_{R_{1}}}{2}(\eta\sigma_{y}\tau_{x}-\sigma_{x}\tau_{y})+M_{s}\sigma_{z}
%+M_{c}
%-\eta\tau_{z}\hbar v_{F}^{2}\frac{\mathcal{A}}{\Omega}
+\lambda_{SOC\perp}\tau_{z}(\eta\sigma_{x}\tau'_{y}+\sigma_{y}\tau'_{x})\\
&+\frac{t_{\perp}}{2}(\tau_{x}\tau'_{x}-\tau_{y}\tau'_{y})
+\frac{t_{w}}{2}(k_{x}\tau_{x}\tau'_{x}+k_{x}\tau_{y}\tau'_{y})
+\mu,
\end{aligned}
\end{equation}
where 
%$P_{x}(t)=k_{x}-\frac{e}{c}A_{x}(t)=k_{x}-\frac{e}{c}A{\rm sin}\Omega t$ with $A$ the scalar potential.
$E_{\perp}$ is the perpendicularly applied electric field, 
$a=3.86$ is the lattice constant,
$\mu$ is the chemical potential,
$\tau'_{x/y}$ denotes the pseudospin of the layers,
$\overline{\Delta}=0.46$ \AA\ is the buckled distance between the upper sublattice and lower sublattice,
$\sigma_{z}$ and $\tau_{z}$ are the spin and sublattice (pseudospin) degrees of freedom, respectively.
$\eta=\pm 1$ for K and K' valley, respectively.
$M_{s}$ is the spin-dependent exchange field. 
$\lambda_{SOC}=3.9$ meV is the strength of intrinsic spin-orbit coupling (SOC) and $\lambda_{R_{2}}=0.7$ meV is the intrinsic Rashba coupling
which is a next-nearest-neightbor (NNN) hopping term and breaks the lattice inversion symmetry.
$\lambda_{R_{1}}$ is the electric field-induced nearest-neighbor (NN) Rashba coupling which is linear with the applied electric field: 
$\lambda_{R_{1}}=0.012E_{\perp}$.

The retarded Green's function in momentum space reads
\begin{equation} 
\begin{aligned}
G(k,\varepsilon+i\eta)=&\frac{1}{\varepsilon+i\eta-H},\\
\end{aligned}
\end{equation}
where $\eta$ is the positive infinitesmall quantity.
Then
the Green's function in real-space can be obtained by the Hamltonian shown in above,
which reads (see Appendix.B for detail derivation)
\begin{equation} 
\begin{aligned}
G(\varepsilon+i\eta,\pm r)
=&G_{1}(\varepsilon,\pm r)\pm{\pmb \tau}\cdot\hat{\bf r}G_{2}(\varepsilon,\pm r)\\
=&\int^{\infty}_{0}\pi I_{0}(iqr)\frac{\varepsilon q}{\varepsilon^{2}-(\hbar v_{F}q)^{2}-f^{2}}dq
\pm{\pmb \tau}\cdot\hat{\bf r}\pi I_{1}(iqr)\int^{\infty}_{0}\frac{q^{2}\hbar v_{F}}{\varepsilon^{2}-(\hbar v_{F}q)^{2}-f^{2}}dq\\
\approx &\int^{\infty}_{0}\pi\frac{e^{iqr}}{\sqrt{2\pi iqr}}\frac{\varepsilon q}{\varepsilon^{2}-(\hbar v_{F}q)^{2}-f^{2}}dq
\pm{\pmb \tau}\cdot\hat{\bf r}
\int^{\infty}_{0}\pi\frac{e^{iqr}}{\sqrt{2\pi iqr}}
\frac{q^{2}\hbar v_{F}}{\varepsilon^{2}-(\hbar v_{F}q)^{2}-f^{2}}dq,
%\int^{\infty}_{0}\pi\frac{e^{iqr}}{\sqrt{2\pi iqr}}\frac{\varepsilon q}{\varepsilon^{2}-(\hbar v_{F}q)^{2}-f^{2}}dq
%+{\pmb \tau}\cdot\hat{\bf r}\int\frac{d^{2}q}{(2\pi)^{2}}\frac{(\pm \hbar v_{F}q{\rm cos}\theta+f)}
%{\varepsilon^{2}-\hbar^{2} v_{F}^{2}q^{2}-f^{2}}e^{ i{\bf q}\cdot{\bf r}},
\end{aligned}
\end{equation}
where we define $G_{1}(\varepsilon,\pm r)$ the pseudospin-independent part and $G_{2}(\varepsilon,\pm r)$ the pseudospin-dependent part,
and we also obtain
\begin{equation} 
\begin{aligned}
G(\varepsilon+i\eta,\pm r)=
%\pi\frac{\sqrt{\frac{2}{\pi}}e^{iqr}-(1+i)\sqrt{qr}{\rm erfi}(\frac{(1+i)\sqrt{qr}}{\sqrt{2}})}{\hbar v_{F}\sqrt{iqr}},
\pi\frac{(\frac{1}{2}+\frac{i}{2})\sqrt{iqr}{\rm erfi}(\frac{(1+i)\sqrt{qr}}{\sqrt{2}})}{\hbar v_{F}\sqrt{q}r^{3/2}},
\end{aligned}
\end{equation}
for the zero-energy state ($\varepsilon=0$),
and its plot is shown in Fig.2 at short distance with different values of $q$.
From Fig.2,
We can see that oscillation behavior is rised by the increasing $q$.
The imaginary part of the Green's functiona in momentum space and the real space are also related to the density of states (DOS)
and local density of states (LDOS), respectively,
by $DOS=\frac{-1}{\pi}{\rm Im}G(\varepsilon+i\eta,q)$ and $LDOS=\frac{-1}{\pi}{\rm Im}G(\varepsilon+i\eta,\pm r)$,
%{Effect of the Rashba splitting on the RKKY interaction in topological-insulator thin films}
%{Probing decoupled edge states in a zigzag phosphorene nanoribbon via RKKY exchange interaction}
and the pole of the Green's function corresponds to the peak of the the DOS and LDOS.

The RKKY interaction between two magnetic impurities which have the local magnetic moment (or the spinor) ${\bf I}_{1}$ and ${\bf I}_{2}$,
respectively, can be described as
%\cite{Wang S X,Zare M}
\begin{equation} 
\begin{aligned}
H_{RKKY}=F_{H}{\bf I}_{1}\cdot{\bf I}_{2}+F_{I}({\bf I}_{1}\cdot\hat{\bf r})({\bf I}_{2}\cdot\hat{\bf r})
+F_{D}({\bf I}_{1}\times{\bf I}_{1})\cdot\hat{\bf r},
\end{aligned}
\end{equation}
where ${\bf I}=(I_{x},I_{y},I_{z})$, $({\bf I}\cdot\hat{\bf r})=(I_{x}{\rm cos}\phi_{r},I_{y}{\rm sin}\phi_{r},I_{z})$
wih ${\rm tan}\phi_{r}=r_{y}/r_{x}$,
and the range functions are
%{Topological phase and edge states dependence of the RKKY interaction in zigzag silicene nanoribbon}
\begin{equation} 
\begin{aligned}
F_{H}=&-J^{2}\frac{2}{\pi}{\rm Im}\int^{\mu}_{-\infty}d\varepsilon\chi_{xx}\\
=&-J^{2}\frac{2}{\pi}{\rm Im}\int^{\mu}_{-\infty}d\varepsilon 
{\rm Tr}[\sigma_{x}G(\varepsilon+i\eta, r)\sigma_{x}G(\varepsilon+i\eta, -r)],\\
F_{I}=&-J^{2}\frac{2}{\pi}{\rm Im}\int^{\mu}_{-\infty}d\varepsilon(\chi_{zz}-\chi_{xx})\\
=&-J^{2}\frac{2}{\pi}{\rm Im}\int^{\mu}_{-\infty}d\varepsilon (
{\rm Tr}[\sigma_{z}G(\varepsilon+i\eta, r)\sigma_{z}G(\varepsilon+i\eta, -r)]
-{\rm Tr}[\sigma_{x}G(\varepsilon+i\eta, r)\sigma_{x}G(\varepsilon+i\eta, -r)]),\\
F_{D}=&-J^{2}\frac{2}{\pi}{\rm Im}\int^{\mu}_{-\infty}d\varepsilon \chi_{xy}\\
=&-J^{2}\frac{2}{\pi}{\rm Im}\int^{\mu}_{-\infty}d\varepsilon 
{\rm Tr}[\sigma_{x}G(\varepsilon+i\eta, r)\sigma_{y}G(\varepsilon+i\eta, -r)],\\
\end{aligned}
\end{equation}
where $J$ is the strength of the $s-d$ interaction (spin exchange interaction) 
between the magnetic impurity and the conduction (itinerant) electrons.
Then by calculating the spin susceptibility tensor (see Appendix.B for detail derivation)
\begin{equation} 
\begin{aligned}
\mathcal{G}_{\alpha\beta}=&\frac{-2}{\pi}{\rm Im}\int^{\mu}_{-\infty}\chi_{\alpha\beta}\\
=&\frac{-2}{\pi}{\rm Im}\int^{\mu}_{-\infty}
{\rm Tr}[\sigma_{\alpha}G(\varepsilon+i\eta, r)\sigma_{\beta}G(\varepsilon+i\eta, -r)],
\end{aligned} 
\end{equation}
the final expression of the the RKKY Hamiltonian
can be obtained as integral over the occupied states
\begin{equation} 
\begin{aligned}
H_{RKKY}^{k}=\frac{-2J^{2}}{\pi}{\rm Im}\int^{\mu}_{-\infty}
\left[\sum_{\alpha\beta}{\bf I}_{1\alpha}
{\rm Tr}[2G_{0}^{2}(\varepsilon+i\eta, r)\delta_{\alpha\beta}-2G_{1}^{2}(\varepsilon+i\eta, r)
(2\delta_{\alpha k}\delta_{\beta k}-\delta_{\alpha\beta})]
{\bf I}_{2\beta}\right].
\end{aligned} 
\end{equation}
We can see that, for the case of $\alpha\neq \beta$ 
%(i.e., the non-diagonal elements of the spin susceptibility tensor),
the RKKY intertaction only contributed by the Heisenberg term and Ising term
due to the absence of the effect of $G_{0}(\varepsilon+i\eta, r)$,
%while for the case of $\alpha=\beta$ and $\alpha\neq k$ or $\beta\neq k$,
and thus the antisymmetry term is missing.
%Specially, at half-filling ($\mu=0$) with electron-hole symmetry,
%the ferromagnetic and the antiferromagnetic type of the RKKT interaction can be found between the magnetic impurities in same
%sublattices and opposite sublattices, respectively,
%which is valid for the bipartite lattice\cite{Saremi S}
%as well as the tripartite lattice\cite{Plumer M L}.

The Eq.(7) can also be written as
\begin{equation} 
\begin{aligned}
G(\varepsilon+i\eta,\pm r)
=&G_{1}(\varepsilon, r)\pm{\pmb \tau}\cdot\hat{\bf r}G_{2}(\varepsilon, r)\\
=&\varepsilon\pi{\rm atanh}\frac{\hbar v_{F}qI_{0}(iqr)}{\hbar(\varepsilon+f)(\varepsilon-f)}\\
&\mp{\pmb \tau}\cdot\hat{\bf r} \frac{1}{2\hbar}(\pi{\rm ln}(f^{2}-\varepsilon^{2}+\hbar^{2}v_{F}^{2}q^{2})I_{1}(iqr)),\\
\end{aligned}
\end{equation}
after integral over the momentum $q$.
%The results at large distance are shown in the Fig.3 where we set the momentum as $q=2$ eV.
We show two terms $G_{1}(\varepsilon, r)$ and $G_{2}(\varepsilon, r)$ in Fig.3,
where we can see that, the
behavior of the retarded real-space Green's function is dominated by the term $G_{1}(\varepsilon,\pm r)$
which oscillates withe distance,
while the term $G_{2}(\varepsilon,\pm r)$ is very small and decrease suddenly at $r=12$.
By substituting it into the Eq.(10),
we can obtain range functions with
%For half-filling case ($\mu=0$), we 
\begin{equation} 
\begin{aligned}
\chi_{H}&=\chi_{xx}=2G^{2}_{1}(\varepsilon, r)-2G^{2}_{2}(\varepsilon, r),\\
\chi_{I}&=\chi_{zz}-\chi_{xx}=%2G^{2}_{1}(\varepsilon, r)+2G^{2}_{2}(\varepsilon,\pm r)-
4G^{2}_{2}(\varepsilon, r),\\
\chi_{D}&=\chi_{xy}=0,
\end{aligned}
\end{equation}
where we set the index $k=x$,
and the results are shown in the Fig.4.

%After some algebra,
%the terms contribute to RKKY interaction can be obtained as
%%{Bulk RKKY signatures of topological phase transition in silicene}
%\begin{equation} 
%\begin{aligned}
%F_{H}=&A+{\rm cos}(2qr)B,\\
%F_{I}=&B+{\rm cos}(2qr)A,\\
%F_{D}=&{\rm sin}(2qr)C,\\
%\end{aligned}
%\end{equation}
%where $A=2\Pi_{\tau_{z}}[\varepsilon-m_{D}]I_{0}(\frac{r\sqrt{m_{D}^{2}-\varepsilon^{2}}}{\hbar v_{F}})$,
%$B=\sum_{\tau_{z}}[\varepsilon-m_{D}]^{2}I^{2}_{0}(\frac{r\sqrt{m_{D}^{2}-\varepsilon^{2}}}{\hbar v_{F}})$,
%$C=\sum_{\tau_{z}}\tau[\varepsilon-m_{D}]^{2}I^{2}_{0}(\frac{r\sqrt{m_{D}^{2}-\varepsilon^{2}}}{\hbar v_{F}})$.
Here we focus on the pseudospin degree of freedom
and consider the model as a two components of the pseudo-spin,
which also has been appeared in the analysis of the RKKY interaction of the Weyl semimetal\cite{Chang2}.
It's valid even in the presence of the Rashba-coupling,
since the Rashba-coupling won't affects the separation of the pseudospin components in pseudospin space
unlike the connecting waveguides\cite{Hafezi M}.

The relaxation of RKKY interaction is $\sim\frac{{\rm sin}(2{\bf k}_{F}r)}{r^{2}}$ for the gapped 2D Dirac system
or the 2DEG in the long-distance limit,
while for the gapless Dirac or Weyl system, it acts as $\sim\frac{{\rm cos}(2{\bf k}_{F}r)}{r^{3}}$\cite{Zare M2,Chang2}.
%{Strongly anisotropic RKKY interaction in monolayer black phosphorus}
%{RKKY interaction of magnetic impurities in Dirac and Weyl semimetals}
Such behavior is similar to the Friedel oscillation as observed in the Dirac or Weyl systems\cite{Wu C H3,XX,Wu C H_3,Chang1}.

For the case of anisotropic dispersion,
the RKKY interaction as well as the DM term and the spin distribution are anisotropic,
expecially for the hole-doped simple.
We pick the black phosphorus as an example,
which with a sizable anisotropic electronic mobility\cite{Fei R}.
The black phosphorus has a highly buckled structure and its edge modes in zigzag direction are quasiflat and nearly 
isolated from the bulk part\cite{Islam S K F},
that provides good platform to observe and measure the RKKY interaction experimentally,
otherwise the measurable RKKY interaction requires the high density of states like the the edge modes\cite{Zare M}.
To deal with the anisotropic RKKY interaction, we at first write the real-space Green's function as
\begin{equation} 
\begin{aligned}
G(\varepsilon, r)
=\frac{1}{4\pi^{2}}\int\int \frac{dq_{x}dq_{y}}{\varepsilon+i\eta-H}e^{iq_{x}r_{x}}e^{iq_{y}r_{y}},
\end{aligned}
\end{equation}
where the low-energy Hamiltonian reads
%{Band topology and the quantum spin Hall effect in bilayer graphene}
%{Universal absorption of two-dimensional materials within k · p method}
%\cite{Prada E}
\begin{equation} 
\begin{aligned}
H=
\begin{pmatrix}[1.5]
(\eta\lambda_{SOC}+\frac{\overline{\Delta}}{2}E_{\perp})\tau_{z}+a_{x}q_{x}^{2}+a_{y}q_{y}^{2}&-\frac{(\eta q_{x}-iq_{y})^{2}}{2m^{*}}\\
-\frac{(\eta q_{x}+iq_{y})^{2}}{2m^{*}}&-(\eta\lambda_{SOC}+\frac{\overline{\Delta}}{2}E_{\perp})\tau_{z}-b_{x}q_{x}^{2}-b_{y}q_{y}^{2},
\end{pmatrix}.
\end{aligned}
\end{equation}
And here $a_{x/y}$ and $b_{x/y}$ are the parameters measuring the degree of anisotropic of black phosphorus
for conduction band and valence band, respectively.
%{Universal absorption of two-dimensional materials within k · p method}
%{Strongly anisotropic RKKY interaction in monolayer black phosphorus}
Then through some algebra,
we can obtain the following spin susceptibility as 
%(see Appendix.C for detail derivation)
\begin{equation} 
\begin{aligned}
\chi_{H}
=&\frac{i}{8\pi^{2}a_{x}b_{x}}\frac{1}{(\sqrt{A}-ik_{x})(\sqrt{A}+k_{x})},\\
\chi_{D}=&0,\\
\chi_{I}
=&(\frac{-i}{4\pi a_{x}}\frac{1}{\sqrt{A}+k_{x}})^{2}+(\frac{-1}{4\pi b_{x}}\frac{1}{\sqrt{A}-ik_{x}})^{2},
%F_{I}=&-J^{2}\frac{2}{\pi}{\rm Im}\int^{\mu}_{-\infty}d\varepsilon (
%{\rm Tr}[\sigma_{z}G(\varepsilon+i\eta, r)\sigma_{z}G(\varepsilon+i\eta, -r)]
%-{\rm Tr}[\sigma_{x}G(\varepsilon+i\eta, r)\sigma_{x}G(\varepsilon+i\eta, -r)]),\\
%F_{D}=&-J^{2}\frac{2}{\pi}{\rm Im}\int^{\mu}_{-\infty}d\varepsilon 
%{\rm Tr}[\sigma_{x}G(\varepsilon+i\eta, r)\sigma_{y}G(\varepsilon+i\eta, -r)].\\
\end{aligned}
\end{equation}
The detail derivation of the above expressions has been presented in the Appendix.C.
Then the Heisenberg, Ising, and DM interaction terms can be obtained by substituting the above expressions into the Eq.(9).

Since for black phosphorus
we have
$a_{x}>a_{y}$ in conduction band and $b_{y}>b_{x}$ in valence band
(here we define the $q_{x}$ in zigzag direction and $q_{y}$ in armchair direction),
we can know that, for electron-doped case ($n$-doped),
the RKKY interaction along zigzag direction is larger than that along the armchair direction,
while for hole-doped case ($p$-doped),
the RKKY interaction along armchair direction is larger than that along the zigzag direction,
which can be proved by the anisotropic dispersion of the black phosphorus\cite{Zare M}.
In fact, for (bilayer) silicene,
the anisotropic dispersion is also possible when under the biaxial strain\cite{Yan J A}
or uniaxial strain\cite{Qin R}.
%then the above procedure is also valid for the 
%calculation of the real-space Green's function.
Base on the DFT result of the Ref.\cite{Elahi M},
we estimate the parameters as:
$a_{x}=0.03$, $a_{y}=0.008$,
$b_{x}=0.005$, $b_{y}=0.038$,
then the factor $\chi$ can be obtained as shown in the Fig.5.

\section{Summary}

We investigate the effect of the non-magnetic and magnetic impurity to the 2D parabolic Dirac system.
At strong screening case within the Thomas-Fermi approximation,
the Thomas-Fermi wave vector is beyong the Fermi surface and it's independent of the carrier density.
%{Boltzmann transport and residual conductivity in bilayer graphene}
as observed by the experiments\cite{Adam S}.
The induced charge density by the charged impurity is also obtained by the linear response theory within the RPA.
%{Quantum criticality of topological phase transitions in three-dimensional interacting electronic systems}
Considering the pseudospin degree of freedom,
we obtained the expression of the RKKY interaction through detail calculations,
the DM term is missing in our system.
The symmetry of the pseudospin space won't be broken even when the effect of Rashba coupling 
is contained.
For the anisotropic dispersion,
the RKKY interaction is also anisotropic and related to the lattice parameters which can be obtained throuh the DFT calculation or
the experiments.
For gapped 2D Dirac system,
the oscillation of the RKKY interaction falls as ${\rm sin}(2k_{F}r)/r^{2}$ at large distance,
that's incontrast to the result of teh gapless Dirac system which falls as ${\rm cos}(2k_{F}r)/r^{3}$,
and the Friedel oscillation of the screened potential shows similar features.
The reason is related to the suppressed backscattering by the Berry phase which is $2\pi$ for the bilayer silicene or graphene\cite{OO}.
Our calculation about the anisotropic RKKY interaction is also illuminating to the study of the anisotropic 3D systems, like the
BiTe$X$ family.

\clearpage

\section{Appendix.A: Induced charge density}

The induced charge density by the charged impurity within RPA reads
\begin{equation} 
\begin{aligned}
\delta n_{r}=&{\rm Ze}\int\frac{d^{2}q}{(2\pi)^{2}}e^{i{\bf q}\cdot{\bf r}}(1-\epsilon(q))\\
=&{\rm Ze}\int\frac{d^{2}q}{(2\pi)^{2}}e^{i{\bf q}\cdot{\bf r}}V_{q}\Pi(q)\\
=&\frac{{\rm Ze}}{4\pi^{2}}\int^{\pi}_{0}d\theta\int^{\infty}_{0}qe^{iqr{\rm cos}\theta}\Pi(q)V_{q}dq,
\end{aligned}
\end{equation}
then for the case of $q\gg m_{D}$,
where $\theta$ is the angle between ${\bf q}$ and ${\bf r}$,
$m_{D}$ is the Dirac-mass which related to the band gap by $\Delta=2m_{D}v_{F}^{2}$,
the static polarization can be approximately obtained as $\Pi(q)=-q(1/4-m_{D}^{2}/q^{2})$\cite{Kotov V N},
thus the induced charge density can be obtained as
\begin{equation} 
\begin{aligned}
\delta n_{r}=&\frac{1}{4\pi^{2}}\frac{{\rm Ze}e^{2}}{2\pi\epsilon\epsilon^{*}}\int^{\infty}_{0}\pi I_{0}(iqr)[-q(\frac{1}{4}-\frac{m_{D}^{2}}{q^{2}})]dq,
\end{aligned}
\end{equation}
where $I_{0}(z)$ is the zeroth-order modified Bessel function of the first kind,
${\rm erfi(z)}$ is the imaginary error function.
%$Ei(z)=-\int^{\infty}_{z}\frac{e^{-x}dx}{x}$ is the exponential integral.
%{http://www.mhtlab.uwaterloo.ca/courses/me755/web_chap4.pdf}
%{Bessel Functions of the First and Second Kind}
By asymptotic approximation,
the above expression can be solved as
\begin{equation} 
\begin{aligned}
\delta n_{r}=
%\frac{{\rm Ze}e^{2}}{2\pi\epsilon\epsilon^{*}}\frac{4m_{D}^{2}qr^{2}Ei(iqr)+e^{iqr}(q+4im_{D}^{2}r)}{8\pi qr^{2}}\bigg|_{\Lambda},
\frac{1}{4\pi^{2}}\frac{{\rm Ze}e^{2}}{2\epsilon\epsilon^{*}}
\frac{\sqrt[4]{-1}\sqrt{\pi}\sqrt{q}(16m_{D}^{2}r^{2}-1){\rm erfi}(\sqrt[4]{-1}\sqrt{qr})+2i\sqrt{r}e^{iqr}(q+8im_{D}^{2}r)}{8\sqrt{2\pi}r^{3/2}\sqrt{iqr}}
\bigg|_{\Lambda},
\end{aligned}
\end{equation}
here $\Lambda$ is the curoff in the ultraviolet limit where the integral convergent.
While for the case of $q\ll m_{D}$,
%where $m_{D}$ is the Dirac-mass which related to the band gap by $\Delta=2m_{D}v_{F}^{2}$,
the static polarization can be approximately obtained as $\Pi(q)=-q^{2}/3\pi m_{D}$\cite{Kotov V N},
then the induced charge density can be obtained as
\begin{equation} 
\begin{aligned}
\delta n_{r}=
%\frac{{\rm Ze}e^{2}}{2\pi\epsilon\epsilon^{*}}\frac{e^{iqr}(qr+i)}{6\pi^{2}mr^{3}}\bigg|_{\Lambda},
\frac{1}{4\pi^{2}}\frac{{\rm Ze}e^{2}}{2\epsilon\epsilon^{*}}
\frac{(-1)^{3/4}3\sqrt{\pi q}{\rm erfi}(\sqrt[4]{-1}\sqrt{qr})+2q\sqrt{r}e^{iqr}(3-2iqr)}{12\sqrt{2}\pi^{3/2}m_{D}r^{5/2}\sqrt{iqr}}\bigg|_{\Lambda},
\end{aligned}
\end{equation}
in asymptotic approximation.
For the more general cases that the $q$ is comparable with the $m_{D}$,
the static polarization function can be written at zero-temperature as
\cite{Wu C H3,XX,Wu C H_3,Tabert C J,Wu C H_4,OO}
%{Dynamical polarization, screening, and plasmons in gapped graphene}
\begin{equation} 
\begin{aligned}
\Pi(q)=-g_{s}g_{v}\frac{2e^{2}\mu}{\epsilon\epsilon^{*}\hbar^{2}v_{F}^{2}}
\left[\frac{ m_{D}}{2\mu}+\frac{\hbar^{2} v_{F}^{2}{\bf q}^{2}-4 m_{D}^{2}}{4\hbar v_{F}{\bf q}\mu}{\rm asin}\sqrt{\frac{\hbar^{2}v_{F}^{2}{\bf q}^{2}}{\hbar^{2}v_{F}^{2}{\bf q}^{2}+4 m_{D}^{2}}}\right]
\end{aligned}
\end{equation}
for $0<\mu< m_{D}$ (intrinsic case), and
\begin{equation} 
\begin{aligned}
\Pi(q)=-g_{s}g_{v}\frac{2e^{2}\mu}{\epsilon_{0}\epsilon \hbar^{2} v_{F}^{2}}
&\left[1-\Theta({\bf q}-2{\bf k}_{F})\right.\\
&\left.\times\left(\frac{\hbar^{2} v_{F}^{2}\sqrt{{\bf q}^{2}-4{\bf k}_{F}^{2}}}{2\hbar v_{F}{\bf q}}-\frac{\hbar^{2}v_{F}^{2}{\bf q}^{2}-4 m_{D}^{2}}{4\mu \hbar v_{F}{\bf q}}{\rm atan}\frac{\hbar v_{F}\sqrt{{\bf q}^{2}-4{\bf k}^{2}_{F}}}{2\mu}\right)\right]
\end{aligned}
\end{equation}
for $\mu> m_{D}$ (extrinsic case).
Then the induced charge density can be obtained by the same procedure.
We discuss the linear Dirac system in above,
while for the 2D parabolic system,
the anisotropic effect can be seen in the effective mass $m^{*}=t_{\perp}/(2 v_{F}^{2})=2\hbar^{2}t_{\perp}/(3 a^{2}t^{2})
=\sqrt{m_{x}m_{y}}$\cite{OO},
note that here the Fermi velocity $v_{F}$ and intralayer/interlayer hopping $t$/$t_{\perp}$ are all anisotropic.
The static polarization can be written as\cite{Low T}
\begin{equation} 
\begin{aligned}
\Pi(q)=
D_{F}^{par}{\rm Re}\left[1-\sqrt{1-\frac{8\mu}{\hbar^{2}}\frac{m^{*2}}{q_{x}^{2}m_{y}+q_{y}^{2}m_{x}}}\right],
\end{aligned}
\end{equation}
then when the effective mass satisfies $m^{*}\sim \frac{2q^{2}\hbar^{2}}{8\sqrt{2}\mu}$
(or when $\mu\rightarrow 0$ in the long-wavelength limit),
the induced charge density of the parabolic system can be obtained as
\begin{equation} 
\begin{aligned}
\delta n_{r}^{par}=
\frac{{\rm Ze}e^{2}}{2\pi\epsilon\epsilon^{*}}
\frac{(\frac{1}{4}+\frac{i}{4})g_{s}m^{*}q^{3/2}\sqrt{r}{\rm erfi}(\frac{(1+i)\sqrt{qr}}{\sqrt{2}})}{\pi\hbar^{2}(iqr)^{3/2}}
\bigg|_{\Lambda},
\end{aligned}
\end{equation}
where $D_{F}^{par}=\frac{g_{s}m^{*}}{2\pi\hbar^{2}}$ is the density of states (DOS) at Fermi level,

\section{Appendix.B: Isotropic}

The retarded Green's function in energy-momentum representation is
\begin{equation} 
\begin{aligned}
G(k,\varepsilon+i\eta)=&\frac{1}{\varepsilon+i\eta-H},\\
\end{aligned}
\end{equation}
where $\eta$ is the positive infinitesmall quantity.
The non-interacting Hamiltonian of bilayer silicene shown in above 
can be simplified as
\begin{equation} 
\begin{aligned}
H=&\frac{\hbar^{2}v_{F}^{2}}{t_{\perp}}
\begin{pmatrix}[1.5]
0&(k_{x}+ik_{y})^{2}\\
(k_{x}-ik_{y})^{2}&0\\
\end{pmatrix}
+\hbar v_{w}
\begin{pmatrix}[1.5]
0&k_{x}+ik_{y}\\
k_{x}-ik_{y}&0\\
\end{pmatrix}\\
&+
(M_{s}\sigma_{z}+\eta\sigma_{z}\lambda_{SOC}+\frac{\overline{\Delta}}{2}E_{\perp})
\begin{pmatrix}[1.5]
1&0\\
0&-1\\
\end{pmatrix}
+
\frac{\hbar^{2}v_{F}^{2}}{t^{2}_{\perp}}\eta\sigma_{z}
\begin{pmatrix}[1.5]
(k_{x}-ik_{y})(k_{x}+ik_{y})&0\\
0&-(k_{x}+ik_{y})(k_{x}-ik_{y})\\
\end{pmatrix},
\end{aligned}
\end{equation}
where we ignore the NN and NNN Rashba-coupling which are quantitatively unimportant.
Then the Green's function in energy-momentum representation is
\begin{equation} 
\begin{aligned}
G(k,\varepsilon+i\eta)=&\frac{1}{\varepsilon+i\eta-(\hbar v_{F}k\cdot{\pmb \tau}+f)}\\
=&\frac{\varepsilon+i\eta+(\hbar v_{F}{\bf k}\cdot{\pmb \tau}+f)}
{(\varepsilon+i\eta)^{2}-(\hbar v_{F}{\bf k}\cdot{\pmb \tau}+f)^{2}}\\
=&\frac{\varepsilon+i\eta+(\hbar v_{F}{\bf k}\cdot{\pmb \tau}+f)}
{(\varepsilon+i\eta)^{2}-\hbar^{2} v_{F}^{2}k^{2}-f^{2}},
\end{aligned}
\end{equation}
where we approximate $(\hbar v_{F}{\bf k}\cdot{\pmb \tau}+f)^{2}\approx(\hbar v_{F}k\cdot{\pmb \tau})^{2}+f^{2}$ for a further calculation
(which is possible when $f\gg \hbar v_{F}k$ or $f\ll \hbar v_{F}k$),
and define $f=f_{1}+f_{2}+f_{3}+f_{4}$,
$f_{1}=(M_{s}+\eta\tau_{z}\lambda_{SOC})\sigma_{z}$,
$f_{2}=\frac{\overline{\Delta}}{2}E_{\perp}\tau_{z}$,
$f_{3}=\frac{t_{\perp}}{2}({\pmb \tau}\cdot{\pmb \tau'}^{*})$,
$f_{4}=\frac{t_{w}}{2}[k_{x}({\pmb \tau}\cdot{\pmb \tau'})])$,
with ${\bf k}=(\eta k_{x},k_{y})$,
${\pmb \tau}=(\eta\tau_{x},\tau_{y})$,
$\sigma=(\sigma_{x},\sigma_{y})$.

In $\eta\rightarrow 0$ limit, we obtain the real-space Green's function
\begin{equation} 
\begin{aligned}
G(\varepsilon+i\eta,\pm r)=&\int\frac{d^{2}q}{(2\pi)^{2}}G(q,\varepsilon)e^{\pm i{\bf q}\cdot{\bf r}}\\
=&\int\frac{d^{2}q}{(2\pi)^{2}}\frac{\varepsilon+(\hbar v_{F}{\bf q}\cdot{\pmb \tau}+f)}
{\varepsilon^{2}-\hbar^{2} v_{F}^{2}q^{2}-f^{2}}e^{\pm i{\bf q}\cdot{\bf r}}\\
=&\int\frac{d^{2}q}{(2\pi)^{2}}\frac{\varepsilon}
{\varepsilon^{2}-\hbar^{2} v_{F}^{2}q^{2}-f^{2}}e^{ i{\bf q}\cdot{\bf r}}
+{\pmb \tau}\cdot\hat{\bf r}\int\frac{d^{2}q}{(2\pi)^{2}}\frac{(\pm \hbar v_{F}q\cdot\hat{\bf r}+f)}
{\varepsilon^{2}-\hbar^{2} v_{F}^{2}q^{2}-f^{2}}e^{ i{\bf q}\cdot{\bf r}}\\
=&\int\frac{d^{2}q}{(2\pi)^{2}}\frac{\varepsilon}
{\varepsilon^{2}-\hbar^{2} v_{F}^{2}q^{2}-f^{2}}e^{ i{\bf q}\cdot{\bf r}}
\pm{\pmb \tau}\cdot\hat{\bf r}\int\frac{d^{2}q}{(2\pi)^{2}}\frac{( \hbar v_{F}q{\rm cos}\theta+f)}
{\varepsilon^{2}-\hbar^{2} v_{F}^{2}q^{2}-f^{2}}e^{ i{\bf q}\cdot{\bf r}}\\
%=&\int^{\pi}_{0}d\theta\int^{\infty}_{0}q G(q,\varepsilon+i\eta)e^{iqr{\rm cos}\theta}dq\\
=&\int^{\infty}_{0}\pi I_{0}(iqr)\frac{\varepsilon q}{\varepsilon^{2}-(\hbar v_{F}q)^{2}-f^{2}}dq
\pm{\pmb \tau}\cdot\hat{\bf r}\int\frac{d^{2}q}{(2\pi)^{2}}\frac{( \hbar v_{F}q{\rm cos}\theta+f)}
{\varepsilon^{2}-\hbar^{2} v_{F}^{2}q^{2}-f^{2}}e^{ i{\bf q}\cdot{\bf r}}\\
\approx &\int^{\infty}_{0}\pi\frac{e^{iqr}}{\sqrt{2\pi iqr}}\frac{\varepsilon q}{\varepsilon^{2}-(\hbar v_{F}q)^{2}-f^{2}}dq
\pm {\pmb \tau}\cdot\hat{\bf r}\int\frac{d^{2}q}{(2\pi)^{2}}\frac{(\hbar v_{F}q{\rm cos}\theta+f)}
{\varepsilon^{2}-\hbar^{2} v_{F}^{2}q^{2}-f^{2}}e^{ i{\bf q}\cdot{\bf r}}\\
=&G_{1}(\varepsilon, r)\pm{\pmb \tau}\cdot\hat{\bf r}G_{2}(\varepsilon, r)
\end{aligned}
\end{equation}
where $\hat{{\bf r}}={\bf r}/r$, and we decompose the momentum ${\bf q}$ into the 
components parallel and perpendiculat to ${\bf r}$.
For convenience, we define the two terms $G_{1}(\varepsilon,\pm r)$ and $G_{2}(\varepsilon,\pm r)$ as shown in above expression.
For a comparation between the $G_{1}(\varepsilon,\pm r)$ and $G_{2}(\varepsilon,\pm r)$,
we firstly note that
\begin{equation} 
\begin{aligned}
\frac{\int e^{iqr{\rm cos}\theta}(\pm\hbar v_{F}q{\rm cos}\theta+f) d\theta}
{\int e^{iqr{\rm cos}\theta} d\theta}
=ir{\rm cos}\theta(\frac{{\rm sec}\theta(\pm\hbar v_{F}-irf)}{r^{2}}-\frac{i(\pm\hbar v_{F}q)}{r})
=\hbar v_{F}q{\rm cos}\theta+f+\frac{i\hbar v_{F}}{r},
\end{aligned}
\end{equation}
we see that there is an imaginary term,
which is consistent with the result of the Weyl semimetal in the presence of TRI and inversion symmetry\cite{Chang2}.

For the massless (with zero Dirac-mass) case, $f=0$,
we obtain
\begin{equation} 
\begin{aligned}
G(\varepsilon,\pm r)
&=G_{1}(\varepsilon, r)\pm{\pmb \tau}\cdot\hat{\bf r}G_{2}(\varepsilon, r)\\
&=\int^{\infty}_{0}\pi\frac{e^{iqr}}{\sqrt{2\pi iqr}}\frac{\varepsilon q}{\varepsilon^{2}-(\hbar v_{F}q)^{2}}dq
\pm{\pmb \tau}\cdot\hat{\bf r}\pi I_{1}(iqr)\int^{\infty}_{0}\frac{q^{2}\hbar v_{F}}{\varepsilon^{2}-(\hbar v_{F}q)^{2}}dq\\
&=\int^{\infty}_{0}\pi\frac{e^{iqr}}{\sqrt{2\pi iqr}}\frac{\varepsilon q}{\varepsilon^{2}-(\hbar v_{F}q)^{2}}dq
\pm{\pmb \tau}\cdot\hat{\bf r}\pi I_{1}(iqr)\int^{\infty}_{0}\frac{q^{2}\hbar v_{F}}{\varepsilon^{2}-(\hbar v_{F}q)^{2}}dq\\
&=\int^{\infty}_{0}\pi\frac{e^{iqr}}{\sqrt{2\pi iqr}}\frac{\varepsilon q}{\varepsilon^{2}-(\hbar v_{F}q)^{2}}dq
\pm{\pmb \tau}\cdot\hat{\bf r}
\int^{\infty}_{0}\pi\frac{e^{iqr}}{\sqrt{2\pi iqr}}
\frac{q^{2}\hbar v_{F}}{\varepsilon^{2}-(\hbar v_{F}q)^{2}}dq\\
&\xlongequal{\varepsilon\rightarrow 0}
%\pi\frac{\sqrt{\frac{2}{\pi}}e^{iqr}-(1+i)\sqrt{qr}{\rm erfi}(\frac{(1+i)\sqrt{qr}}{\sqrt{2}})}{\hbar v_{F}\sqrt{iqr}}
\pm\pi\frac{(\frac{1}{2}+\frac{i}{2})\sqrt{iqr}{\rm erfi}(\frac{(1+i)\sqrt{qr}}{\sqrt{2}})}{\hbar v_{F}\sqrt{q}r^{3/2}},
\end{aligned}
\end{equation}
%where ${\rm erf}(z)=\frac{2}{\sqrt{\pi}}\int^{z}_{0}e^{-t^{2}}dt$ is the error function,
where $I_{1}(z)$ is the firts-order modified Bessel function of the first kind,
and it has $I_{1}(z)=I_{0}(z)$ in the first-order of the asymptotic approximation.
The last equality of the above expression is valid in the limit of $\varepsilon\rightarrow 0$,
i.e, the zero-energy state.

As shown in the main text, the trace taking over the spin and pseudospin degrees of freedom reads
\begin{equation} 
\begin{aligned}
{\rm Tr}[\sigma_{\alpha}G(\varepsilon+i\eta, r)\sigma_{\beta}G(\varepsilon+i\eta, -r)]=
&G_{1}(\varepsilon+i\eta, r)G_{1}(\varepsilon+i\eta, r){\rm Tr}(\sigma_{\alpha}\sigma_{0}\sigma_{\beta}\sigma_{0})\\
&-G_{2}(\varepsilon+i\eta, r)G_{2}(\varepsilon+i\eta, r){\rm Tr}(\sigma_{\alpha}{\pmb \tau}\sigma_{\beta}{\pmb \tau})\\
&+G_{2}(\varepsilon+i\eta, r)G_{1}(\varepsilon+i\eta, r){\rm Tr}(\sigma_{\alpha}{\pmb \tau}\sigma_{\beta}\sigma_{0})\\
&-G_{1}(\varepsilon+i\eta, r)G_{2}(\varepsilon+i\eta, r){\rm Tr}(\sigma_{\alpha}\sigma_{0}\sigma_{\beta}{\pmb \tau})\\
=&2G_{1}^{2}(\varepsilon+i\eta, r)\delta_{\alpha\beta}-2G_{2}^{2}(\varepsilon+i\eta, r)(2\delta_{\alpha k}\delta_{\beta k}-\delta_{\alpha\beta}),
\end{aligned}
\end{equation}
%{RKKY interaction in three-dimensional electron gases with linear spin-orbit coupling}
where $\tau_{k}\ (k=x,y,z)$ denotes pseodospin degree of freedom.
We can see that,
due to the existence of the pseudospin degree of freedom,
the spin suscepetibility is not the one reported in Ref.\cite{Zare M} which considers only the degree of freedom of real spin.
Then the RKKY Hamiltonian can be obtained as integral over the occupied states (from valence band to the Fermi level)
\begin{equation} 
\begin{aligned}
H_{RKKY}^{k}=\frac{-2J^{2}}{\pi}{\rm Im}\int^{\mu}_{-\infty}
\left[\sum_{\alpha\beta}{\bf I}_{1\alpha}
{\rm Tr}[2G_{1}^{2}(\varepsilon+i\eta, r)\delta_{\alpha\beta}-2G_{2}^{2}(\varepsilon+i\eta, r)
(2\delta_{\alpha k}\delta_{\beta k}-\delta_{\alpha\beta})]
{\bf I}_{2\beta}\right].
\end{aligned} 
\end{equation}
Thus we can see that, for the case of $\alpha\neq \beta$ (i.e., the non-diagonal elements of the spin susceptibility tensor),
the RKKY intertaction only contributed by the Heisenberg term and Ising term
due to the absence of the effect of $G_{0}(\varepsilon+i\eta, r)$,
%while for the case of $\alpha=\beta$ and $\alpha\neq k$ or $\beta\neq k$,
and thus the antisymmetry term is missing.
Specially, at half-filling ($\mu=0$) with electron-hole symmetry,
the ferromagnetic and the antiferromagnetic type of the RKKT interaction 
between the site-impurities can be found between the magnetic impurities in same
sublattices and opposite sublattices, respectively,
which is valid for the bipartite lattice\cite{Saremi S}
as well as the tripartite lattice\cite{Plumer M L}.
Otherwise,
the ferromegnati and antiferromagnetic type of the RKKY interactions (not the site-impurity)
induced by the perpendicular magnetic moment may related to the strength of the spin susceptibility\cite{Zare M}
of the Heisenberg type and the Ising type.

\section{Appendix.C: Anisotropic parabolic system}

For the anisotropic 2D Dirac system, like the (bilayer) black phosphorus,
we write down the real-space Green's function as
%{Strongly anisotropic RKKY interaction in monolayer black phosphorus}
%\cite{Prada E}
\begin{equation} 
\begin{aligned}
G(\varepsilon, r)
=\frac{1}{4\pi^{2}}\int\int \frac{dq_{x}dq_{y}}{\varepsilon+i\eta-H}e^{iq_{x}r_{x}}e^{iq_{y}r_{y}},
\end{aligned}
\end{equation}
where the low-energy Hamiltonian in two-band model (for $t_{\perp}\gg\varepsilon$) reads
%{Band topology and the quantum spin Hall effect in bilayer graphene}
%{Universal absorption of two-dimensional materials within k · p method}
%\cite{Prada E}
\begin{equation} 
\begin{aligned}
H=
\begin{pmatrix}[1.5]
(\eta\lambda_{SOC}+\frac{\overline{\Delta}}{2}E_{\perp})\tau_{z}+a_{x}q_{x}^{2}+a_{y}q_{y}^{2}&-\frac{(\eta q_{x}-iq_{y})^{2}}{2m^{*}}\\
-\frac{(\eta q_{x}+iq_{y})^{2}}{2m^{*}}&-(\eta\lambda_{SOC}+\frac{\overline{\Delta}}{2}E_{\perp})\tau_{z}-b_{x}q_{x}^{2}-b_{y}q_{y}^{2},
\end{pmatrix},
\end{aligned}
\end{equation}
where the diagonal elements refer to the intraband coupling while the non-diagonal elements
refer to the interband coupling,
$a_{x/y}$ and $b_{x/y}$ terms are refer to the remote-band coupling (indirecte intraband coupling)
of the conduction band and valence band respectively,
and here $a_{x}\neq a_{y}$, $b_{x}\neq b_{y}$
for the anisotropic system,
which can be easily observed in, e.g., balck phosphorus and the 
then the eigenenergy can be obtained by solving the above expression,
\begin{equation} 
\begin{aligned}
\varepsilon_{\pm}=
%\frac{1}{m^{*2}}\left(
%\frac{\overline{\Delta}}{2}E_{\perp}m^{*2}\tau_{z}\pm\sqrt{m^{*4}\lambda_{SOC}^{2}+4m^{*2}{\rm cos}^{4}\phi+8m^{*2}{\rm cos}^{2}\phi
%{\rm sin}^{2}\phi+4m^{*2}{\rm sin}^{4}\phi}
%\right),
&(1/(m^{2}))(\frac{\overline{\Delta}}{2}E_{\perp} m^{2} \tau_{z} + 2 a_{x} m^{2} {\rm cos}^{2}\phi + 2 b_{x} m^{2} {\rm cos}^{2}\phi + 
  2 a_{y} m^{2} {\rm sin}^{2}\phi + 
  2 b_{y} m^{2} {\rm sin}^{2}\phi- \sqrt{F}),\\
F=&
( m^{4} \lambda_{SOC}^{2} \tau^{2} + 
     4 a_{x} \eta m^{4} \lambda_{SOC} \tau_{z} {\rm cos}^{2}\phi - \\
&     4 (-b_{x}) \eta m^{4} \lambda_{SOC} \tau_{z} {\rm cos}^{2}\phi + 4 \eta^{4} m^{2} {\rm cos}^{4}\phi + 
     4 a_{x}^{2} m^{4} {\rm cos}^{4}\phi - 8 a_{x} (-b_{x}) m^{4} {\rm cos}^{4}\phi + \\
&     4 (-b_{x})^{2} m^{4} {\rm cos}^{4}\phi + 4 a_{y} \eta m^{4} \lambda_{SOC} \tau_{z} {\rm sin}^{2}\phi - 
     4 (-b_{y}) \eta m^{4} \lambda_{SOC} \tau_{z} {\rm sin}^{2}\phi - 
     8  (-1)m^{2} {\rm cos}^{2}\phi {\rm sin}^{2}\phi + \\
&     8 a_{x} a_{y} m^{4} {\rm cos}^{2}\phi {\rm sin}^{2}\phi - 
     8 a_{y} (-b_{x}) m^{4} {\rm cos}^{2}\phi {\rm sin}^{2}\phi - \\
&     8 a_{x} (-b_{y}) m^{4} {\rm cos}^{2}\phi {\rm sin}^{2}\phi + 
     8 (-b_{x}) (-b_{y}) m^{4} {\rm cos}^{2}\phi {\rm sin}^{2}\phi + 4  m^{2} {\rm sin}^{4}\phi + \\
&     4 a_{y}^{2} m^{4} {\rm sin}^{4}\phi - 8 a_{y} (-b_{y}) m^{4} {\rm sin}^{4}\phi + 
     4 (-b_{y})^{2} m^{4} {\rm sin}^{4}\phi ).
\end{aligned}
\end{equation}
where we define $q_{x}=q{\rm cos}\phi$ and $q_{y}=q{\rm sin}\phi$.
Thus the real-space Green's function is
\begin{equation} 
\begin{aligned}
G^{\pm}(\varepsilon, r)
=\frac{1}{4\pi^{2}}\int\int\frac{dq_{x}dq_{y}}{\varepsilon+i\eta-[
\pm (\eta\lambda_{SOC}+\frac{\overline{\Delta}}{2}E_{\perp})\tau_{z}\pm \xi_{x}q_{x}^{2}\pm \xi_{y}q_{y}^{2}]
}e^{iq_{x}r_{x}}e^{iq_{y}r_{y}},
\end{aligned}
\end{equation}
where $\xi=a,b$ for $G^{+}(\varepsilon, r)$ and $G^{-}(\varepsilon, r)$, respectively.
Through some straightforward calculation, for the conduction band sector (in above matrix),
we have
\begin{equation} 
\begin{aligned}
G^{+}(\varepsilon, r)
=&\frac{1}{4\pi^{2}}\int dq_{y}e^{iq_{y}r_{y}}
\int\frac{dq_{x}}{\varepsilon+i\eta-[
(\eta\lambda_{SOC}+\frac{\overline{\Delta}}{2}E_{\perp})\tau_{z}+ a_{x}q_{x}^{2}+ a_{y}q_{y}^{2}]
}e^{iq_{x}r_{x}}\\
=&\frac{1}{4\pi^{2}}\int dq_{y}e^{iq_{y}r_{y}}
\frac{1}{a_{x}}\int^{\infty}_{0}\frac{e^{iq_{x}r_{x}}}{A-q_{x}^{2}}dq_{x}
\end{aligned}
\end{equation}
%{Strongly anisotropic RKKY interaction in monolayer black phosphorus}
where we define $A=\frac{\varepsilon+i\eta-(\eta\lambda_{SOC}+\frac{\overline{\Delta}}{2}E_{\perp})\tau_{z}-a_{y}q_{y}^{2}}{a_{x}}$.
Then we use the residues calculation to deal with the integral
\begin{equation} 
\begin{aligned}
S=\frac{1}{a_{x}}\int^{\infty}_{-\infty}\frac{e^{ik_{x}r_{x}}}{A-k_{x}^{2}}dk_{x}.
\end{aligned}
\end{equation}
Since the poles of the function within integral $S$ is $k_{x}^{1,2}=\pm\sqrt{A}$,
and it satisfies
\begin{equation} 
\begin{aligned}
{\rm Res}[\frac{e^{ik_{x}r_{x}}}{A-k_{x}^{2}},\pm\sqrt{A}]=
\frac{1}{(m-1)!}\lim_{k_{x}\rightarrow k_{x}^{1,2}}\frac{d^{m-1}}{dk_{x}^{m-1}}\{(k_{x}-\pm\sqrt{A})^{m}\frac{e^{ik_{x}r_{x}}}{A-k_{x}^{2}}\},
\end{aligned}
\end{equation}
where $m$ is the order of the poles and here $m\rightarrow \infty$,
thus 
\begin{equation} 
\begin{aligned}
\lim_{k_{x}\rightarrow m}\frac{e^{ik_{x}r_{x}}}{A-k_{x}^{2}}=\lim_{k_{x}\rightarrow \infty}\frac{e^{ik_{x}r_{x}}}{A-k_{x}^{2}}=0.
\end{aligned}
\end{equation}
By restrict the contour in the upper half-plane,
we have
\begin{equation} 
\begin{aligned}
S=\frac{1}{a_{x}}(-2\pi i)[\frac{e^{ik_{x}r_{x}}}{\sqrt{A}+k_{x}}],
\end{aligned}
\end{equation}
thus 
\begin{equation} 
\begin{aligned}
G^{+}(\varepsilon+i\eta, r_{x})
=&\frac{1}{4\pi^{2}a_{x}}\int^{\infty}_{0}\frac{e^{iq_{x}r_{x}}e^{iq_{y}r_{y}}}{A-q_{x}^{2}}dq_{x}\\
=&\frac{1}{4\pi^{2}a_{x}}(-\pi i)[\frac{e^{iq_{x}r_{x}}e^{iq_{y}r_{y}}}{\sqrt{A}+q_{x}}],
\end{aligned}
\end{equation}
while the $y$-direction component can be obtained through the same procedure,
For the valence band sector,
similarly,
we have
\begin{equation} 
\begin{aligned}
S'=&\frac{1}{b_{x}}\int^{\infty}_{-\infty}\frac{e^{ik_{x}r_{x}}}{A'+k_{x}^{2}}dk_{x}\\
=&\frac{1}{b_{x}}(\pi i\frac{e^{ik_{x}r_{x}}}{A'+k_{x}^{2}})(k_{x}-k_{x}^{1})\\
=&\frac{1}{b_{x}}(\pi i\frac{e^{ik_{x}r_{x}}}{A'+k_{x}^{2}})(k_{x}-i\sqrt{A'})\\
=&\frac{1}{b_{x}}(-\pi \frac{e^{ik_{x}r_{x}}}{\sqrt{A'}-ik_{x}})\\.
\end{aligned}
\end{equation}
where $A'=\frac{\varepsilon+i\eta-(-\eta\lambda_{SOC}-\frac{\overline{\Delta}}{2}E_{\perp})\tau_{z}+b_{y}q_{y}^{2}}{b_{x}}$,
and thus
\begin{equation} 
\begin{aligned}
G^{-}(\varepsilon+i\eta, r_{x})
=&\frac{1}{4\pi^{2}b_{x}}\int^{\infty}_{0}\frac{e^{iq_{x}r_{x}}e^{iq_{y}r_{y}}}{A+k_{x}^{2}}dk_{x}\\
=&\frac{1}{4\pi^{2}b_{x}}(-\pi)[\frac{e^{ik_{x}r_{x}}e^{ik_{y}r_{y}}}{\sqrt{A}-ik_{x}}].
\end{aligned}
\end{equation}
For electron-doped case ($n$-doped),
the RKKY interaction along armchair direction is larger than that along the zigzag direction,
while for hole-doped case ($p$-doped),
the RKKY interaction along zigzag direction is larger than that along the armchair direction,
which can be proved by the anisotropic dispersion of the black phosphorus\cite{Zare M}.
In fact, for (bilayer) silicene,
the anisotropic dispersion is also possible when under the biaxial strain\cite{Yan J A}
or uniaxial strain\cite{Qin R},
then the above procedure is also valid for the 
calculation of the real-space Green's function.

Then we write the full real-space Green's function as
\begin{equation} 
\begin{aligned}
G^{\pm}(\varepsilon, r_{x})
\begin{pmatrix}[1.5]
G^{+}(\varepsilon, r_{x}) &0\\
0&G^{-}(\varepsilon, r_{x})
\end{pmatrix},
\end{aligned}
\end{equation}
thus the spin susceptibility are
\begin{equation} 
\begin{aligned}
\chi_{xx}
=&{\rm Tr}[\sigma_{x}G(\varepsilon+i\eta, r)\sigma_{x}G(\varepsilon+i\eta, -r)]\\
=&G^{-}(\varepsilon, r_{x})G^{+}(\varepsilon, -r_{x})+G^{+}(\varepsilon, r_{x})G^{-}(\varepsilon, -r_{x})\\
=&\frac{1}{4\pi^{2}b_{x}}(-\pi)[\frac{e^{ik_{x}r_{x}}e^{ik_{y}r_{y}}}{\sqrt{A}-ik_{x}}]
\frac{1}{4\pi^{2}a_{x}}(-\pi i)[\frac{e^{-ik_{x}r_{x}}e^{-ik_{y}r_{y}}}{\sqrt{A}+k_{x}}]\\
&+
\frac{1}{4\pi^{2}a_{x}}(-\pi i)[\frac{e^{ik_{x}r_{x}}e^{ik_{y}r_{y}}}{\sqrt{A}+k_{x}}]
\frac{1}{4\pi^{2}b_{x}}(-\pi)[\frac{e^{-ik_{x}r_{x}}e^{-ik_{y}r_{y}}}{\sqrt{A}-ik_{x}}]\\
=&\frac{i}{8\pi^{2}a_{x}b_{x}}\frac{1}{(\sqrt{A}-ik_{x})(\sqrt{A}+k_{x})};\\
\chi_{xy}
=&{\rm Tr}[\sigma_{x}G(\varepsilon+i\eta, r)\sigma_{y}G(\varepsilon+i\eta, -r)]\\
=&-i[G^{+}(\varepsilon, r_{x})G^{-}(\varepsilon, -r_{x})-G^{-}(\varepsilon, r_{x})G^{+}(\varepsilon, -r_{x})]\\
=&-i[\frac{1}{4\pi^{2}a_{x}}(-\pi i)[\frac{e^{ik_{x}r_{x}}e^{ik_{y}r_{y}}}{\sqrt{A}+k_{x}}]
\frac{1}{4\pi^{2}b_{x}}(-\pi)[\frac{e^{-ik_{x}r_{x}}e^{-ik_{y}r_{y}}}{\sqrt{A}-ik_{x}}]-\\&
\frac{1}{4\pi^{2}b_{x}}(-\pi)[\frac{e^{ik_{x}r_{x}}e^{ik_{y}r_{y}}}{\sqrt{A}-ik_{x}}]
\frac{1}{4\pi^{2}a_{x}}(-\pi i)[\frac{e^{-ik_{x}r_{x}}e^{-ik_{y}r_{y}}}{\sqrt{A}+k_{x}}]]\\
=&0;\\
\chi_{zz}
=&{\rm Tr}[\sigma_{z}G(\varepsilon+i\eta, r)\sigma_{z}G(\varepsilon+i\eta, -r)]\\
=&G^{+}(\varepsilon, r_{x})G^{+}(\varepsilon, -r_{x})+G^{-}(\varepsilon, r_{x})G^{-}(\varepsilon, -r_{x})\\
=&\frac{1}{4\pi^{2}a_{x}}(-\pi i)[\frac{e^{ik_{x}r_{x}}e^{ik_{y}r_{y}}}{\sqrt{A}+k_{x}}]
\frac{1}{4\pi^{2}a_{x}}(-\pi i)[\frac{e^{-ik_{x}r_{x}}e^{-ik_{y}r_{y}}}{\sqrt{A}+k_{x}}]+\\&
\frac{1}{4\pi^{2}b_{x}}(-\pi)[\frac{e^{ik_{x}r_{x}}e^{ik_{y}r_{y}}}{\sqrt{A}-ik_{x}}]
\frac{1}{4\pi^{2}b_{x}}(-\pi)[\frac{e^{-ik_{x}r_{x}}e^{-ik_{y}r_{y}}}{\sqrt{A}-ik_{x}}]\\
=&(\frac{-i}{4\pi a_{x}}\frac{1}{\sqrt{A}+k_{x}})^{2}+(\frac{-1}{4\pi b_{x}}\frac{1}{\sqrt{A}-ik_{x}})^{2},
%F_{I}=&-J^{2}\frac{2}{\pi}{\rm Im}\int^{\mu}_{-\infty}d\varepsilon (
%{\rm Tr}[\sigma_{z}G(\varepsilon+i\eta, r)\sigma_{z}G(\varepsilon+i\eta, -r)]
%-{\rm Tr}[\sigma_{x}G(\varepsilon+i\eta, r)\sigma_{x}G(\varepsilon+i\eta, -r)]),\\
%F_{D}=&-J^{2}\frac{2}{\pi}{\rm Im}\int^{\mu}_{-\infty}d\varepsilon 
%{\rm Tr}[\sigma_{x}G(\varepsilon+i\eta, r)\sigma_{y}G(\varepsilon+i\eta, -r)],\\
\end{aligned}
\end{equation}
then the Heisenberg, Ising, and DM interaction terms can be obtained by substituting the above expressions into the Eq.(9).
Here we note that, for a alternative form of the above Green's function,
it can also be written as
\begin{equation} 
\begin{aligned}
G^{k}(\varepsilon, r_{x})
\begin{pmatrix}[1.5]
G^{+}(\varepsilon, r_{x}) &0\\
0&G^{-}(\varepsilon, r_{x})
\end{pmatrix}
\end{aligned}
\end{equation}
even in the presence of the Rashba coupling (here $k$ follows the definition in Appdensix.B),
since that, although the Rashba coupling mixing the up- and down-spin components,
it won't affect the separation of the pseudospin components.
%{Strongly anisotropic RKKY interaction in monolayer black phosphorus}

\end{large}
\renewcommand\refname{References}

\clearpage

\begin{large}

Fig.1
\begin{figure}[!ht]
\subfigure{
\begin{minipage}[t]{0.5\textwidth}
\centering
\includegraphics[width=1\linewidth]{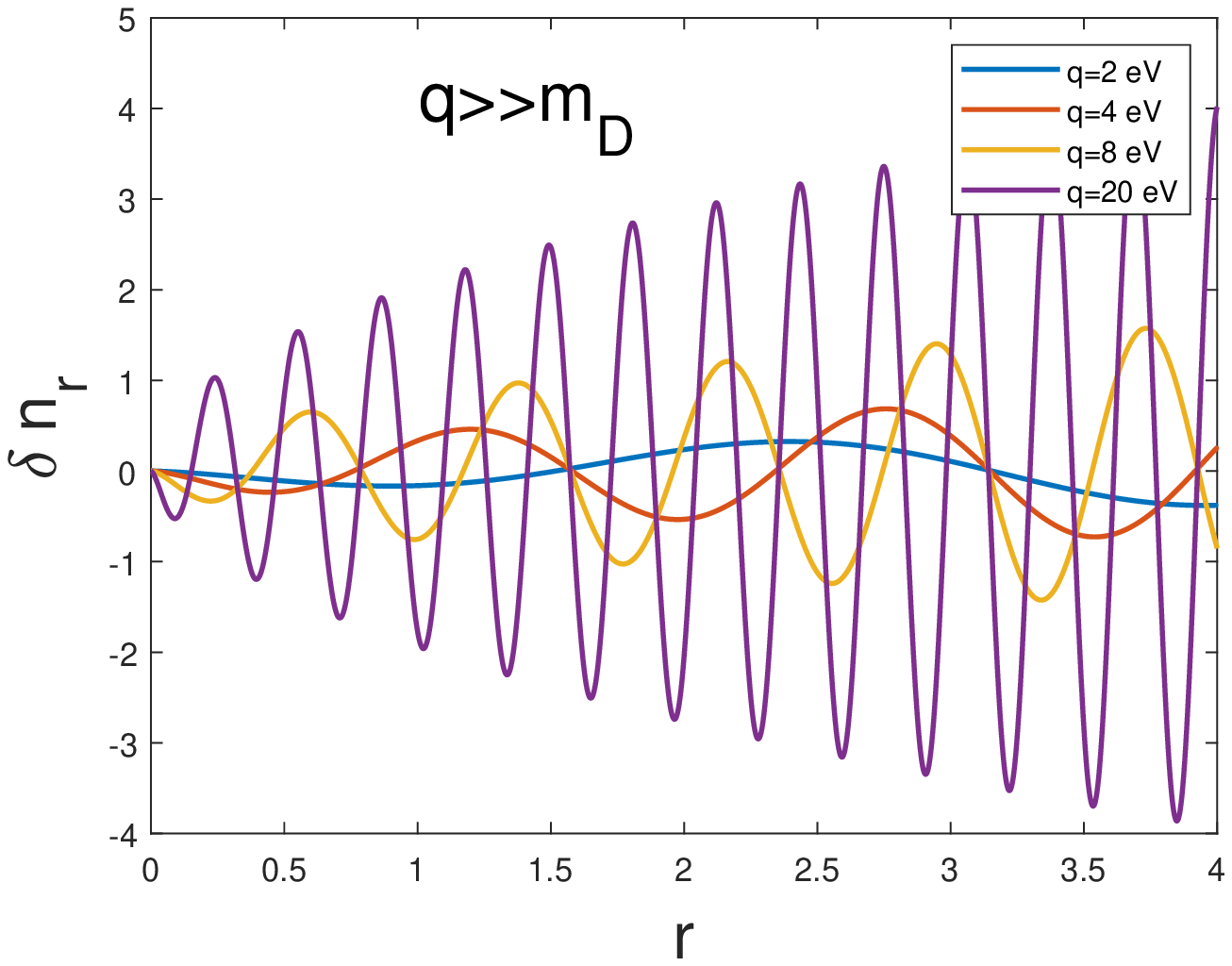}
\label{fig:side:a}
\end{minipage}
}\\
\subfigure{
\begin{minipage}[t]{0.55\textwidth}
\centering
\includegraphics[width=0.9\linewidth]{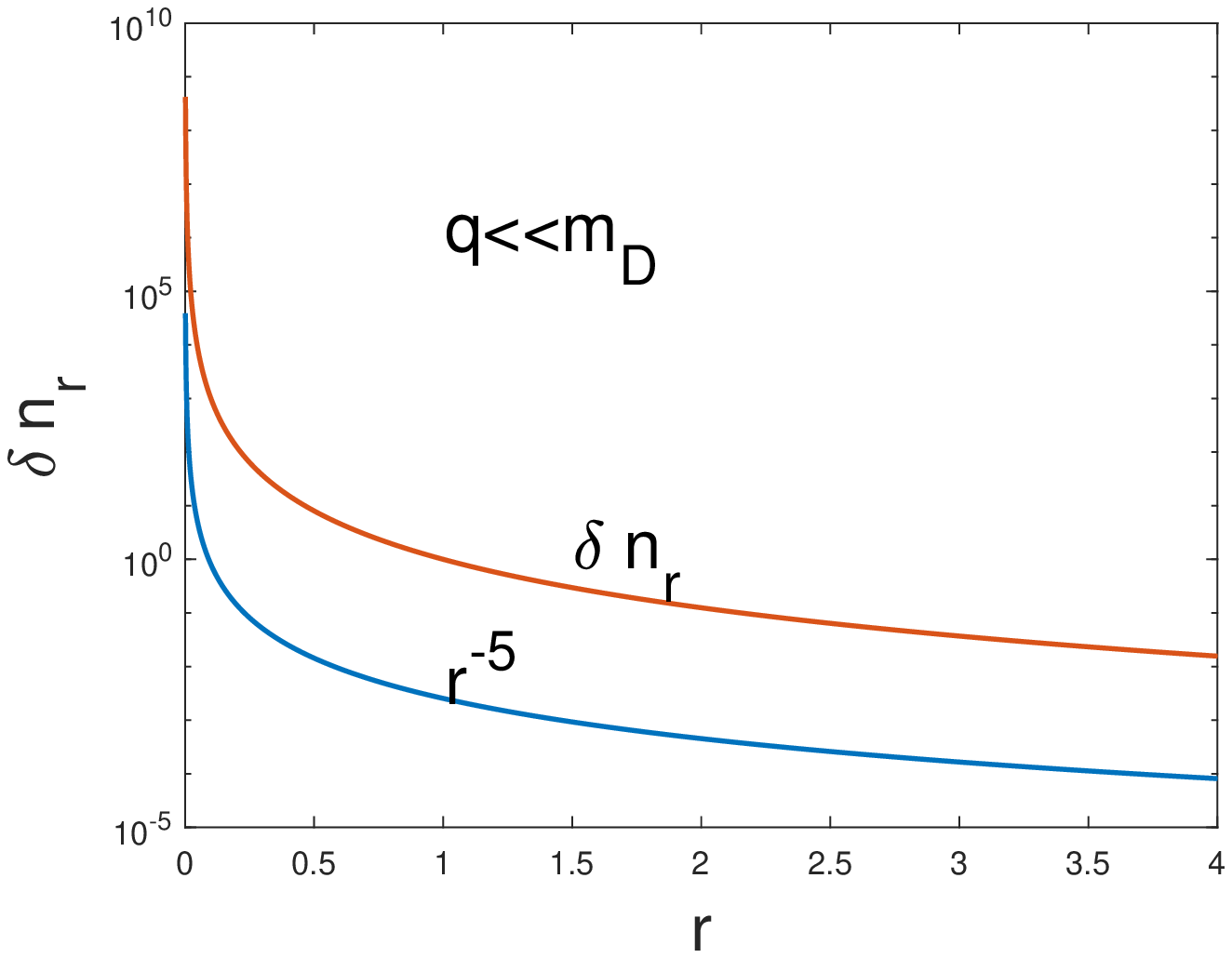}
\label{fig:side:b}
\end{minipage}
}
%{Probing the topological phase transition via density oscillations in silicene and germanene}
\caption{(Color online) The induced charge density by the charged impurity.
The upper panel shows the results in the case of $q\gg m_{D}$, where we set $m_{D}=0.02$ eV here.
The lower panel shows the resultin the case of $q\ll m_{D}$, where we set $q=0.02$ eV and $m_{D}=2$ eV here.
}
\end{figure}
\clearpage
Fig.2
\begin{figure}[!ht]
   \centering
 \centering
   \begin{center}
     \includegraphics*[width=0.6\linewidth]{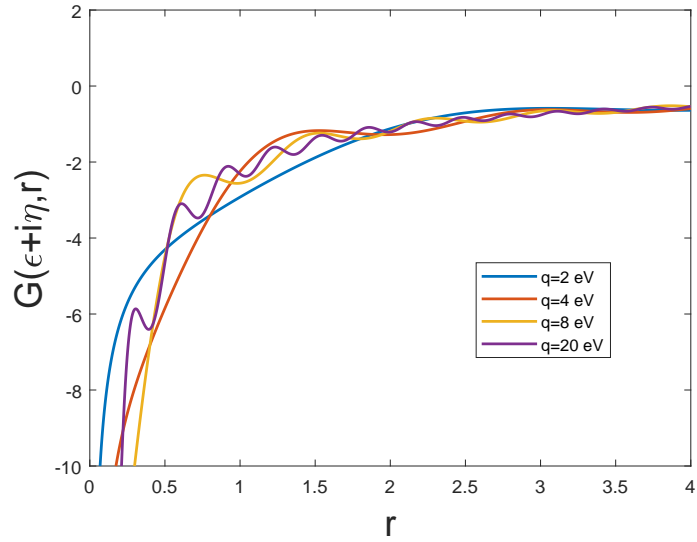}
\caption{(Color online) The approximated result of the retarded real-space Green's function in zero-energy state ($\varepsilon\rightarrow 0$).
We set a list of momentum as indicated.
}
   \end{center}
\end{figure}

Fig.3
\begin{figure}[!ht]
   \centering
 \centering
   \begin{center}
     \includegraphics*[width=1\linewidth]{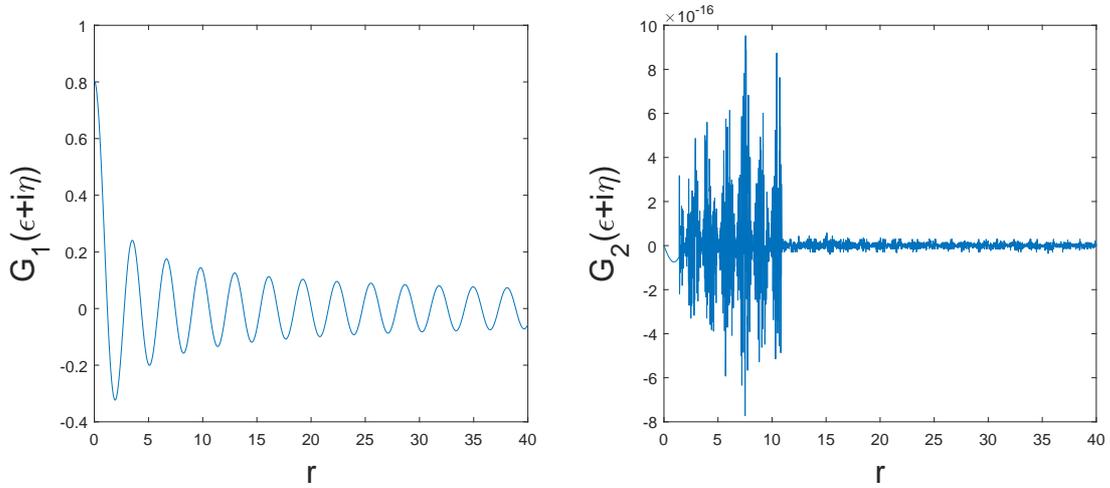}
\caption{(Color online) Large distance behavior of the retarded real-space Green's function 
$G_{1}(\varepsilon, r)$ (left panel) and $G_{2}(\varepsilon, r)$ (right panel).
We set $q=2$ eV, $f=0.02$ eV, $\varepsilon=0.5$ eV here.
}
   \end{center}
\end{figure}
\clearpage
Fig.4
\begin{figure}[!ht]
   \centering
 \centering
   \begin{center}
     \includegraphics*[width=1\linewidth]{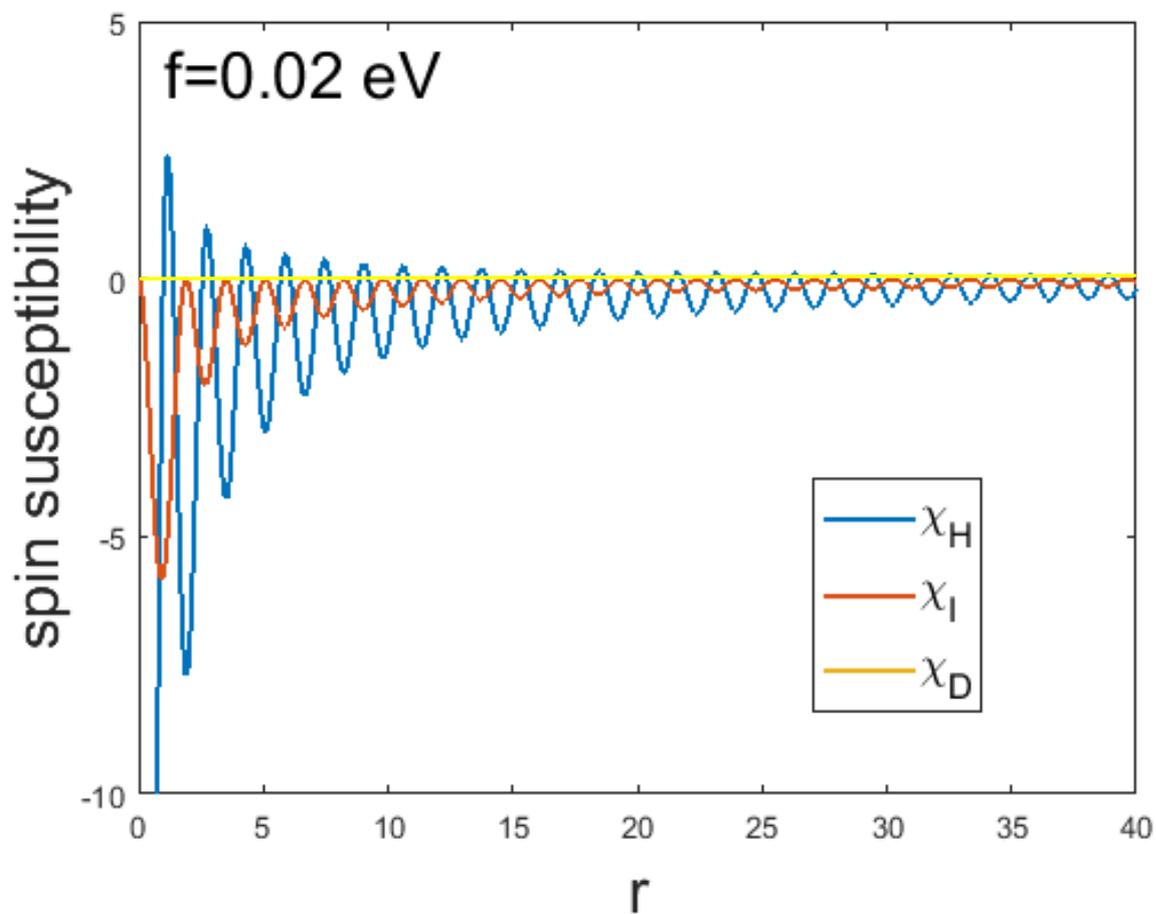}
\caption{(Color online) Spin susceptibility of the Heisenberg-type, Ising-type, and DM-type
for isotropic dispersion.
The indices are setted as $\alpha=\beta$ and thus the DM term vanishes
and the RKKY interaction is contributed by only the Heisenberg term and Ising term.
The $f$-term (see Appendix.A) is setted as 0.02 eV here.
}
   \end{center}
\end{figure}
Fig.5
\begin{figure}[!ht]
   \centering
 \centering
   \begin{center}
     \includegraphics*[width=1\linewidth]{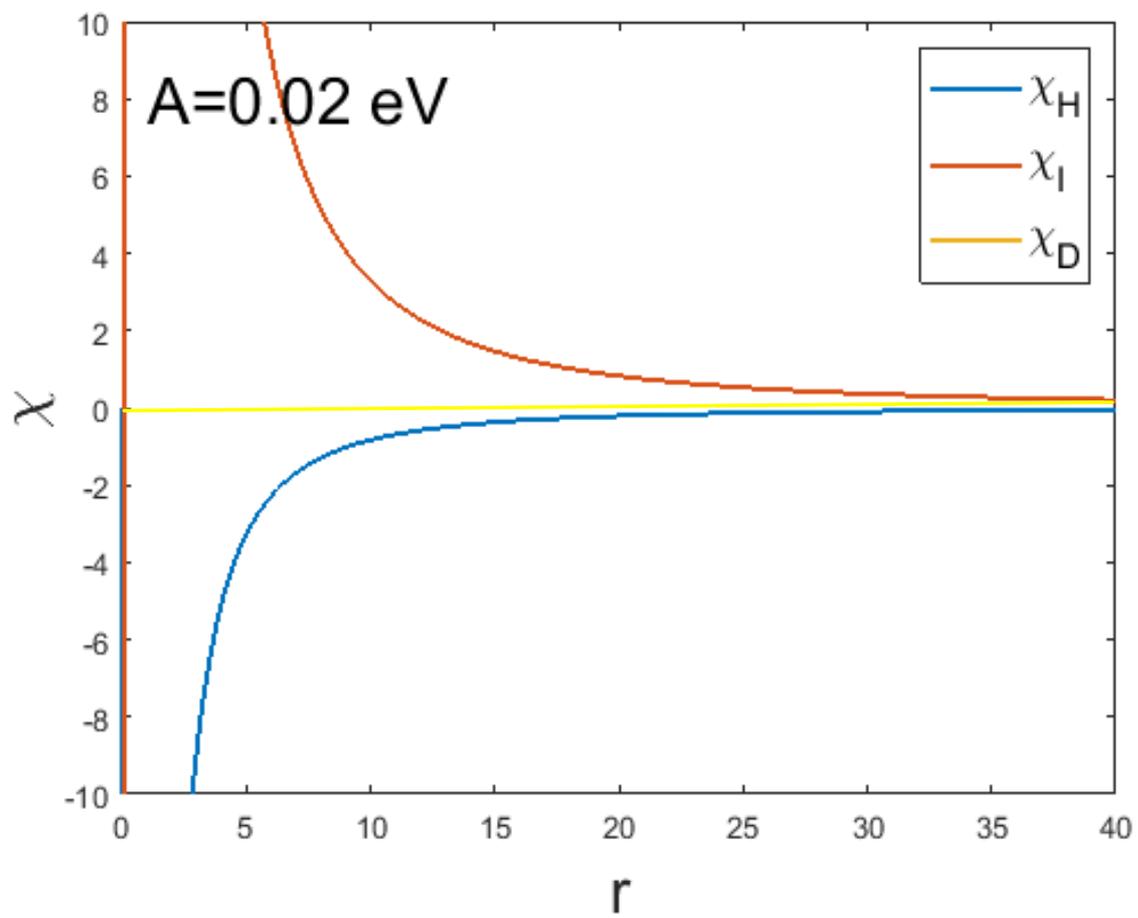}
\caption{(Color online) $\chi$ of the Heisenberg-type, Ising-type, and DM-type
in anisotropic 2D black phosphorus.
We set $A=0.02$ eV here (see Appendix.C).
}
   \end{center}
\end{figure}

\end{large}

\end{document}